\documentclass[aps,amsmath,twocolumn,notitlepage,amssymb,prb,longbibliography]{revtex4-1}

\usepackage{graphicx}
\usepackage{dcolumn}
\usepackage{bm}
\usepackage[colorlinks=true,citecolor=red,urlcolor=blue]{hyperref}
\usepackage{wasysym}
\usepackage{bbold}
\usepackage{stmaryrd}
\usepackage{verbatim}
\usepackage{subfigure}

\begin{document}
\title{Quantum Paramagnet and Frustrated Quantum Criticality \\in a Spin-One Diamond Lattice Antiferromagnet}
\author{Gang Chen$^{1,2}$}
\email{gangchen.physics@gmail.com}
\date{\today}
\affiliation{$^1$State Key Laboratory of Surface Physics, Department of Physics,
Center for Field Theory \& Particle Physics,
Fudan University, Shanghai, 200433, China}
\affiliation{$^2$Collaborative Innovation Center of Advanced Microstructures, 
Nanjing, 210093, China}

\begin{abstract}
Motivated by the proposal of {\it topological quantum paramagnet} 
in the diamond lattice antiferromagnet NiRh$_2$O$_4$, we propose a minimal 
model to describe the magnetic interaction and properties of the diamond 
material with the spin-one local moments. Our model includes the 
first and second neighbor Heisenberg interactions as well as a local 
single-ion spin anisotropy that is allowed by the spin-one nature of 
the local moment and the tetragonal symmetry of the system. 
We point out that there exists a quantum phase transition from a 
{\it trivial quantum paramagnet} when the single-ion spin anisotropy is 
dominant to the magnetic ordered states when the exchange is dominant.  
Due to the frustrated spin interaction, the magnetic excitation in the  
quantum paramagnetic state supports {\it extensively degenerate band minima} 
in the spectra. As the system approaches the transition, extensively 
degenerate bosonic modes become critical at the criticality, giving rise to 
unusual magnetic properties. Our phase diagram and experimental predictions for 
different phases provide a guildline for the identification of the ground 
state for NiRh$_2$O$_4$. Although our results are fundamentally different 
from the proposal of topological quantum paramagnet, 
it represents interesting possibilities for spin-one diamond lattice 
antiferromagnets. 
\end{abstract}

\maketitle

\emph{Introduction.}---The recent theoretical proposal of symmetry protected 
topological (SPT) ordered states has sparked a wide interest in the theoretical 
community~\cite{PhysRevB.80.155131,PhysRevB.81.064439,PhysRevB.83.035107,PhysRevB.85.075125,PhysRevB.87.155114,senthil2015symmetry,PhysRevX.3.011016,PhysRevB.87.174412,PhysRevB.87.235122,PhysRevB.84.235128,chen2012symmetry,song2016topological,PhysRevB.88.035131,PhysRevB.92.125111,PhysRevB.86.125119,chen2014symmetry,PhysRevB.91.134404,PhysRevB.91.184404,PhysRevX.6.041006,PhysRevX.4.041049,geraedts2014exact,PhysRevB.93.205157,PhysRevX.5.021029,PhysRevB.87.195128,PhysRevB.89.045127}. 
The well-known topological insulator, 
that was proposed and discovered 
earlier, is a {\it non-interacting} fermion SPT protected by time reversal symmetry~\cite{hasan2010colloquium,qi2011topological}. 
In contrast, the SPTs in bosonic systems must be stabilized 
by the interactions~\cite{chen2012symmetry}. 
The spin degrees of freedom with exchange interactions seem to 
be a natural candidate for realizing the boson SPTs~\cite{PhysRevB.84.235128}. 
In fact, the Haldane spin-one chain is a 1D boson SPT and 
is protected by the SO(3) spin rotational 
symmetry~\cite{HALDANE1983464,PhysRevB.80.155131,PhysRevB.81.064439}. 
The realization of boson SPTs in high dimensions is still missing. 
It was suggested that, the spin-one diamond lattice antiferromagnet 
with frustrated spin interactions may host a topological quantum 
paramagnet that is a spin analogue of topological insulator and 
protected by time reversal symmetry~\cite{PhysRevB.91.195131}. 
Quite recently, a diamond lattice antiferromagnet NiRh$_2$O$_4$ 
with Ni$^{2+}$ spin-one local moments was proposed to fit into 
the early suggestion~\cite{McQueenunPub}.

\begin{figure}[hb]
{\includegraphics[width=7cm]{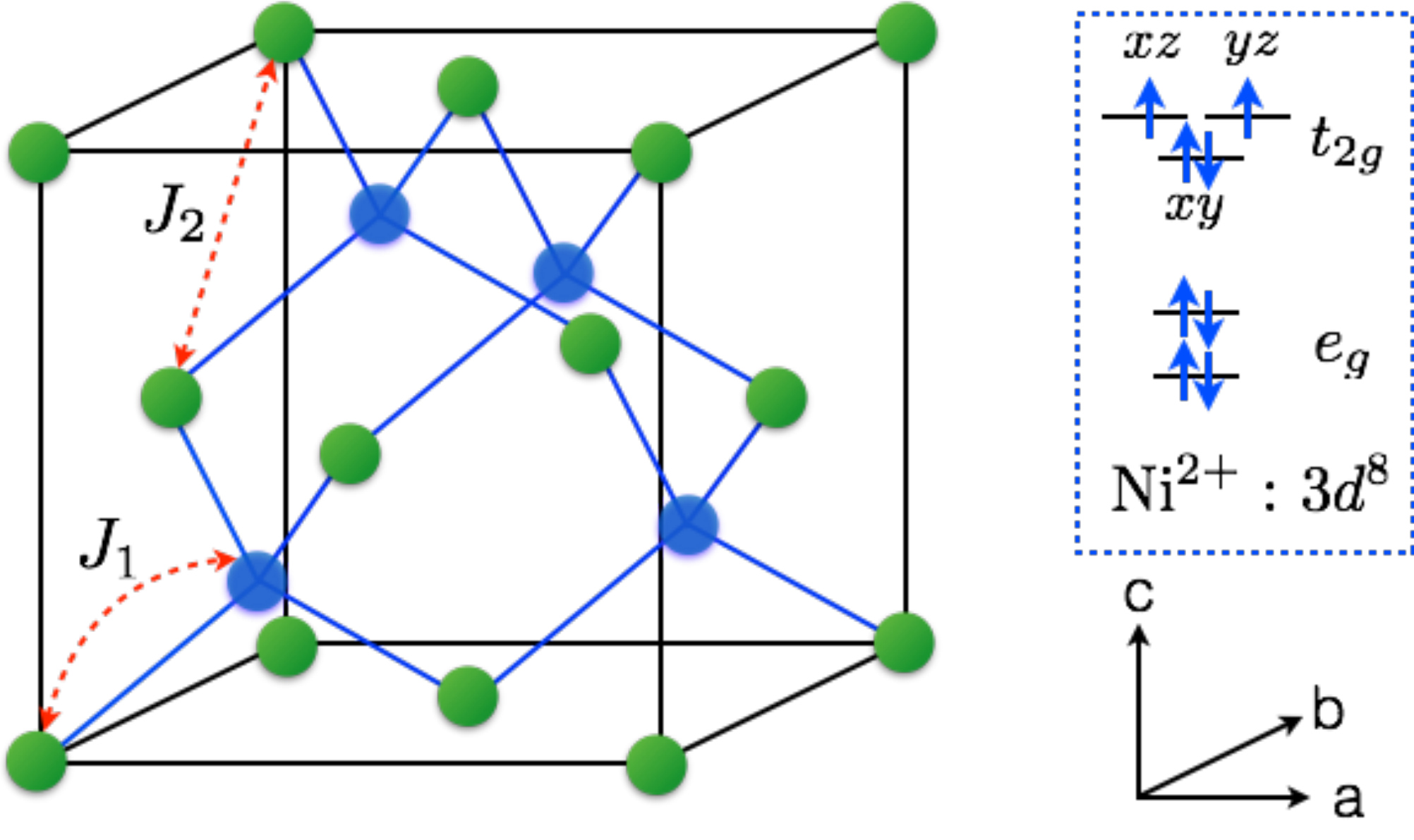}}
\caption{(Color online.) The diamond lattice with the $J_1$ and $J_2$ interactions. 
Due to the tetragonal symmetry of the lattice, 
the $a$ and $b$ directions are not equivalent to the $c$ direction. 
The Ni$^{2+}$ ion is in a tetrahedral environment, so the $e_g$ orbitals
are lower in energy than the $t_{2g}$ levels. The 
tetragonal distortion further splits the two $e_g$ orbitals and the 
three $t_{2g}$ orbitals. But the degeneracy of the $xz$ and $yz$ orbitals
remains intact under the tetragonal distortion. To avoid the orbital 
degree of freedom, we here place the $xz$ and $yz$ orbitals above the $xy$ 
orbitals~\cite{Supple}.
}
\label{fig1}
\end{figure}

NiRh$_2$O$_4$ is a tetragonal spinel and experiences a structural phase 
transition from cubic to tetragonal at ${T = 380}$K~\cite{JPSJ.19.423,McQueenunPub,McQueen}. 
As we show in Fig.~\ref{fig1}, the magnetic ion 
Ni$^{2+}$ has a $3d^8$ electron configuration, forming a spin ${S=1}$ local 
moment and occupying the tetrahedral diamond lattice site. 
No signature of magnetic order was observed down to 0.1K in the 
magnetic susceptibility and specific heat measurements. 
Although this might fulfill the basic requirement of the absence of 
magnetic order in a topological quantum paramagnet, alternative state, 
that is distinct from topological quantum paramagnet, 
may also provide a consistent experimental prediction with 
the current experiments. In this Rapid Communication, 
we propose a minimal spin model for 
spin-one diamond lattice with tetragonal distortion and 
study the full phase diagram and the phase transition of our model. 
We do not find the presence of 
the topological quantum paramagnet in our phase diagram. Instead, due 
to the strong spin frustration, the ordered state in our phase diagram 
can be easily destabilized and converted into a trivial quantum 
paramagnet by a moderate single-ion spin anisotropy. We predict that 
this seemingly trivial quantum paramagnetic state in a large parameter 
regime supports gapped magnetic excitation that develops {\it extensively 
degenerate band minima} in the spectrum. As the quantum paramagnet 
approaches the phase transition to the proximate ordered state, the 
extensively degenerate low-energy modes become gapless and are responsible 
for the unusual magnetic properties such as the 
{\it linear-T heat capacity} at low temperatures in the vicinity 
of the transition. In the proximate ordered phases, we further show that 
the spin spiral orders are actually induced by quantum fluctuations via 
quantum order by disorder.

\emph{The microscopic spin model.}---We here propose the following microscopic spin model
that describes the interaction between the spin-1 local moments with the tetragonal symmetry,
\begin{equation}
H =  J_1^{} \sum_{\langle {\boldsymbol r} {\boldsymbol r}' \rangle} 
{\boldsymbol S}_{\boldsymbol r}^{} 
\cdot {\boldsymbol S}_{{\boldsymbol r}'}^{}
+ J_2^{} \sum_{\langle\langle  {\boldsymbol r} {\boldsymbol r}'\rangle\rangle}  
{\boldsymbol S}_{\boldsymbol r}^{} 
\cdot {\boldsymbol S}_{{\boldsymbol r}'}^{}
+ D_z^{} \sum_{\boldsymbol r} (S_{\boldsymbol r}^z)^2 ,
\end{equation}
where $J_1$ and $J_2$ are the first neighbor and second neighbor 
Heisenberg exchange interactions, respectively. 
Although the tetragonal lattice symmetry allows inequivalent bonds~\cite{McQueen}, 
in this minimal model we assume all the bonds are equivalent.  
Since the diamond lattice is a bipartite lattice, the first neighbor 
$J_1$ interaction alone is unfrustrated, and would favor a simple 
N\'eel state if $J_1$ is antiferromagnetic. The second neighbor 
interaction $J_2$ is an interaction within each FCC sublattice of 
the diamond lattice. Due to the large numbers of second neighbor bonds, 
the $J_2$ interaction would cause a spin frustration even when it is 
small compared to $J_1$. 
Moreover, an additional single-ion spin anisotropy is further introduced 
on top of the spin exchange interactions, and is not 
included in the model in Ref.~\onlinecite{McQueen}. 
The spin anisotropy is naturally allowed by the tetragonal lattice symmetry 
and is the only term occuring for a spin-one local moment like the Ni$^{2+}$ 
ion. Previous classical treatment of the $J_1$-$J_2$ 
spin model on a diamond lattice and the analysis of thermal fluctuation 
have led to the interesting discovery of the spiral spin 
liquid~\cite{bergman2007order,gao2016spiral,PhysRevB.78.144417,PhysRevB.84.064438}. 
A quantum treatment of $J_1$-$J_2$ model used an exotic $\text{SP(N)}$ parton construction
for the spins~\cite{PhysRevLett.101.047201} and again worked in the 
ordered regime. In our context, we will 
largely treat spins and interactions quantum mechanically with a more 
conventional means that is appropriate for the $J_1$-$J_2$-$D_z$ model.

Due to this single-ion spin anisotropy, the magnetic susceptibilities along 
different directions should reveal such spin anisotropy. In particular, we carry
out the high temperature series expansion and find that the Curie-Weiss temperatures
for the magnetic field parallel and normal to the $z$ direction are given as~\cite{Supple}
\begin{eqnarray}
\Theta_{\text{CW}}^{z}  &=& - \frac{D_z}{3} - \frac{S(S+1)}{3} (z_1 J_1 + z_2 J_2), \\
\Theta_{\text{CW}}^{\perp} &=& + \frac{D_z}{6} - \frac{S(S+1)}{3} (z_1 J_1 + z_2 J_2), 
\end{eqnarray}
where ${z_1 =4}$ and ${z_2 =12}$ are the numbers of first neighbor and 
second neighbor bonds, respectively. The above prediction can be used to 
extract the single-ion spin anisotropy. Note for a powder sample, 
the Curie-Weiss temperature is ${\Theta_{\text{CW}}^{\text{Powder}} 
= -\frac{S(S+1)}{3} (z_1 J_1 + z_2 J_2)}$ 
and is thus independent of the spin anisotropy.

\begin{figure}[t]
{\includegraphics[width=7cm]{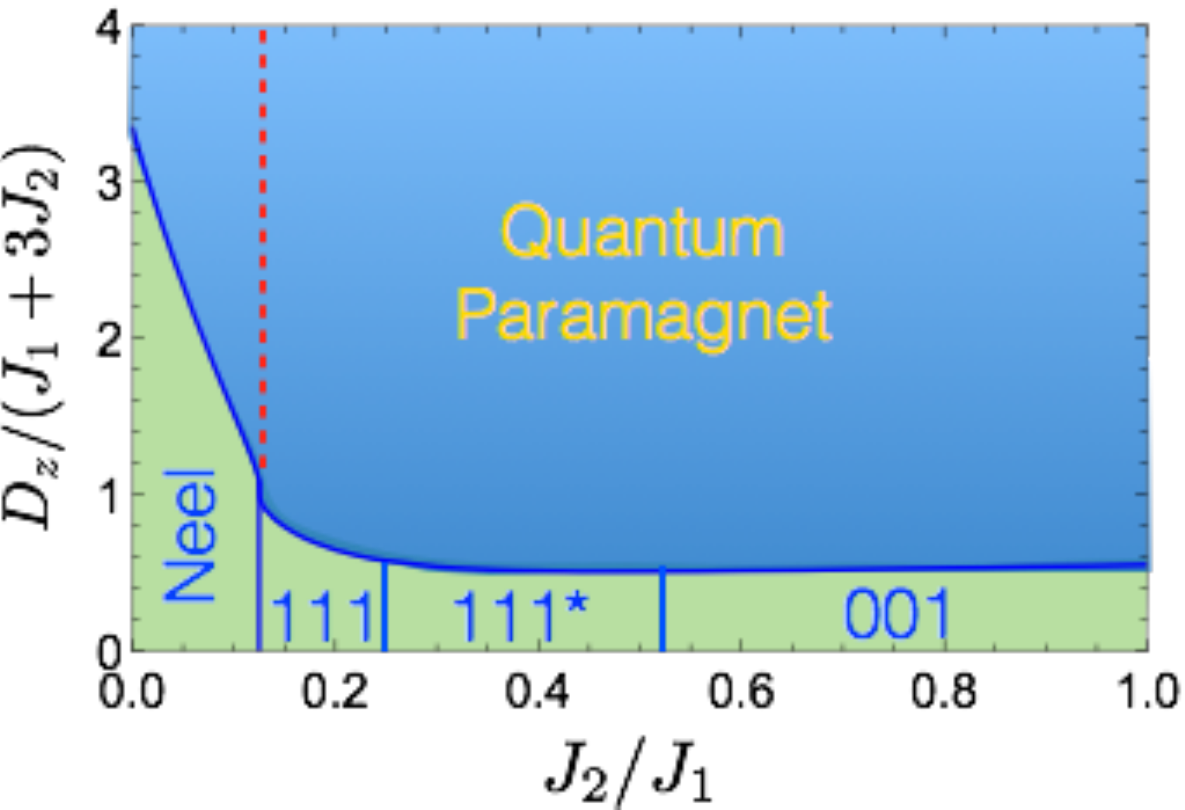}}
\caption{(Color online.) The phase diagram of the $J_1$-$J_2$-$D_z$ model.
Because the powder sample Curie-Weiss temperature ${\Theta_{\text{CW}}^{\text{Powder}} 
= -8(J_1+3J_2)/3}$, we set the energy unit of the spin anisotropy $D_z$ to 
${J_1+3J_2}$ in the plot. The transition from the quantum paramagnet to the 
ordered regions is continuous at the mean-field theory.  
On the left of the (red) dashed line, the band mininum of the magnetic excition 
is unique and appears at $\Gamma$ point. On the right side, the band minima form
a degenerate surface in the reciprocal space. Please refer the main text for 
details. 
}
\label{fig2}
\end{figure}

\emph{Quantum paramagnet and phase diagram.}---To obtain the full phase diagram of 
the $J_1$-$J_2$-$D_z$ model, we start from the parameter regime where the single-ion 
spin anisotropy is dominant. We consider the easy-plane anisotropy with ${D_z > 0}$, 
since the easy-axis spin anisotropy would stabilize the N\'eel state and enlarge its 
parameter regime. In the large and positive $D_z$ limit, the ground state is a 
trivial quantum paramagnet with ${S^z = 0}$ on every site,
$|\Psi \rangle = \prod_{\boldsymbol r} | S^z_{\boldsymbol r} = 0 \rangle$.
For this simple state, there is no magnetic order and all the spin excitations 
are fully gapped. Since the global U(1) spin rotational symmetry around the $z$ 
direction is preserved, the magnetic susceptibility at zero temperature for the 
field along the $z$ direction is zero with ${\chi_{z}^{} (T=0) =0 }$. However, 
if the field is applied in the $xy$ plane, the spin rotational symmetry is broken 
by the in-plane field and the magnetic susceptibility is a constant with 
\begin{eqnarray}
\chi_{\perp}^{} (T=0) =  
\frac{2\mu_0 (g \mu^{}_{\text B})^2}{D_z + 2 (z_1 J_1 + z_2 J_2)},
\end{eqnarray} 
where $g$ is the Lande factor. Again, this result
is a consequence of the single-ion anisotropy and 
can be used to detect the quantum paramagnetic state.

As we turn on the exchange interaction, the spin excitation would 
develop dispersion in the momentum space. With a sufficient exchange 
interaction, we expect the minimum of the dispersion to touch the zero energy 
that would lead to magnetic orderings. To describe the magnetic ordering transition
out of the quantum paramagnetic phase, we substitute the spin operators 
with the rotor variables such that~\cite{PhysRevLett.109.016402} 
\begin{eqnarray}
S^z_{\boldsymbol r} = n_{\boldsymbol r}^{}, \quad\quad
S^{\pm}_{\boldsymbol r} = \sqrt{2} e^{\pm i \phi_{\boldsymbol r}^{}},
\end{eqnarray}
where $\phi^{}_{\boldsymbol r}$ is a $2\pi$-periodic phase variable and 
$n_{\boldsymbol r}$ is integer-valued. This substitution has enlarged 
the physical Hilbert space by allowing $S^z$ or $n$ to take 
the values beyond $0$ and $\pm 1$. We, however, do not expect this 
approximation to cause significant effects since the the non-physical
values of $n^{}_{\boldsymbol r}$ has been energetically suppressed by 
the large single-ion spin anisotropy. Moreover, the substitition 
preserves the global U(1) spin rotational symmetry around the $z$ direction
of the original spin model. Finally, to preserve the spin commutation 
relation, we impose the commutation for $\phi^{}_{\boldsymbol r}$ 
and $n^{}_{\boldsymbol r}$ with 
$[\phi_{\boldsymbol r}, n_{{\boldsymbol r}'}] 
= i \delta_{ {\boldsymbol r}{\boldsymbol r}' }$.
 
With the rotor variables, the $J_1$-$J_2$-$D_z$ spin model 
takes the form 
\begin{eqnarray}
H &=& 
 \sum_{\langle {\boldsymbol r} {\boldsymbol r}' \rangle} J_1^{}
 [2\cos (\phi_{\boldsymbol r}^{} - \phi_{{\boldsymbol r}'}^{}) 
+ n_{\boldsymbol r}^{} n_{{\boldsymbol r}'}^{}   ]
\nonumber \\
&+&  \sum_{\langle\langle {\boldsymbol r} {\boldsymbol r}' \rangle\rangle} J_2^{} 
 [ 2\cos (\phi_{\boldsymbol r}^{} - \phi_{{\boldsymbol r}'}^{}) 
+ n_{\boldsymbol r} n_{{\boldsymbol r}'}   ] 
\nonumber \\
&+&   \sum_{\boldsymbol r}  D_z^{} n^2_{\boldsymbol r}
.
\end{eqnarray}
From the symmetry point of view, the above model has the same symmetry
as a standard boson Hubbard model except having an extra inter-site 
boson interaction. To make this analogy a bit further, 
the quantum paramagnetic state is analogous to 
a boson Mott insulator with ${n_{\boldsymbol r} = 0}$ at every site,
and the proximate magnetic order is like a superfluid of bosons. 
Despite the seemingly similarity, we show below the 
intrinsic spin frustration brings rather interesting dispersion of 
magnetic excitation in the quantum paramagnet and thus 
leads to unusual properties at the analogous ``superfluid-Mott'' 
transition~\cite{PhysRevB.40.546}.

The primary operators that are responsible for the magnetic transition 
out of the quantum paramagnet are the $S^{\pm}_{\boldsymbol r}$ spin 
operators that create the gapped spin excitations in the quantum paramagnet
but take finite values in the ordered states. We here carry out the 
coherent state path integral and integrate out the number operator 
$n_{\boldsymbol r}^{}$. The resulting partition function is
\begin{eqnarray}
Z = \int \mathcal{D} \Phi_{\boldsymbol r}^{}  
\mathcal{D} \lambda_{\boldsymbol r}^{}
\, \text{exp} \big[ {- {\mathcal S} }
-\mathbb{i} \sum_{\boldsymbol r} \lambda_{\boldsymbol r}^{} 
(|\Phi_{\boldsymbol r}^{}|^2 -1) 
\big],
\label{part}
\end{eqnarray}
where the effective action for the rotor variable is 
\begin{eqnarray}
{\mathcal S} &=& \int d\tau \sum_{{\boldsymbol k} \in \text{BZ}} 
(2D_z \mathbb{1}_{2\times 2}^{} + \mathcal{J}_{\boldsymbol k}^{})_{ij}^{-1} 
\partial_{\tau}^{} \Phi^{\dagger}_{i,{\boldsymbol k}} 
\partial_{\tau}^{} \Phi^{}_{j,{\boldsymbol k}}
\nonumber \\
&& 
+ \sum_{\langle {\boldsymbol r}{\boldsymbol r}' \rangle} 
J_1^{} \Phi^{\dagger}_{\boldsymbol r} \Phi^{}_{{\boldsymbol r}'}
+ 
\sum_{\langle\langle {\boldsymbol r}{\boldsymbol r}' \rangle\rangle} 
J_2^{} \Phi^{\dagger}_{\boldsymbol r} \Phi^{}_{{\boldsymbol r}'},
\end{eqnarray}
where we have introduced the variable ${\Phi_{\boldsymbol r}^{} \equiv
e^{i\phi_{\boldsymbol r}^{}}}$. To impose the unimodular condition for 
$\Phi_{\boldsymbol r}^{}$, we have introduced a Lagrange multiplier
$\lambda_{\boldsymbol r}$ on each site to impose the unimodular condition
$|{{\Phi_{\boldsymbol r}^{}}| = 1}$ in Eq.~\eqref{part}. 
To solve for the dispersion of the excitation, we take a saddle point
approximation and choose a uniform mean-field ansatz such that 
${ \mathbb{i} \lambda_{\boldsymbol r}  \equiv \beta \Delta (T)}$ 
where ${\beta = (k_{\text B} T)^{-1}}$. 
We integrate out the $\Phi_{\boldsymbol r}$ field and obtain 
the saddle-point equation for $\Delta(T)$ in the quantum 
paramagnetic phase 
\begin{eqnarray}
\sum_{i=1,2} \sum_{{\boldsymbol k} \in \text{BZ}} 
\frac{2D_z + \xi_{i,{\boldsymbol k}}}
{\omega_{i,{\boldsymbol k}}}
\coth (\frac{\beta \omega_{i,{\boldsymbol k}}}{2} ) 
= 2, 
\label{saddle}
\end{eqnarray}
where $\omega_{1,{\boldsymbol k}}$ and $\omega_{2,{\boldsymbol k}}$   
are the two modes of the magnetic excitations in the paramagnetic 
phase and are given by
\begin{eqnarray}
\omega_{i,{\boldsymbol k}}
= \big[ (4D_z + 2 \xi_{i,{\boldsymbol k}})
(\Delta (T) + \xi_{i,{\boldsymbol k}})
 \big]^{\frac{1}{2}},
 \label{excitation}
\end{eqnarray}
and $\xi_{1, {\boldsymbol k}}$ and $\xi_{2, {\boldsymbol k}}$
are the two eigenvalues of the exchange matrix 
${\mathcal J}_{\boldsymbol k}$~\cite{Supple}. 
As one decreases the single-ion spin anisotropy, the gap 
of the magnetic excitation decreases steadily. At the 
transition, the gap is closed and induces the  
magnetic order, and this phase transition is continuous
within this treatment. In the phase diagram that is depicted 
in Fig.~\ref{fig2}, the phase boundary between the quantum 
paramagnet and the magnetic order is then determined by 
examining the gap of the excitations in Eq.~\eqref{excitation}. 
In Fig.~\ref{fig2}, the ordered region of the phase diagram 
is further splited into several sub-regions with distinct
magnetic orders from the quantum order by disorder effect. 
This will be explained below soon.

\begin{figure}[t]
{\includegraphics[width=8cm]{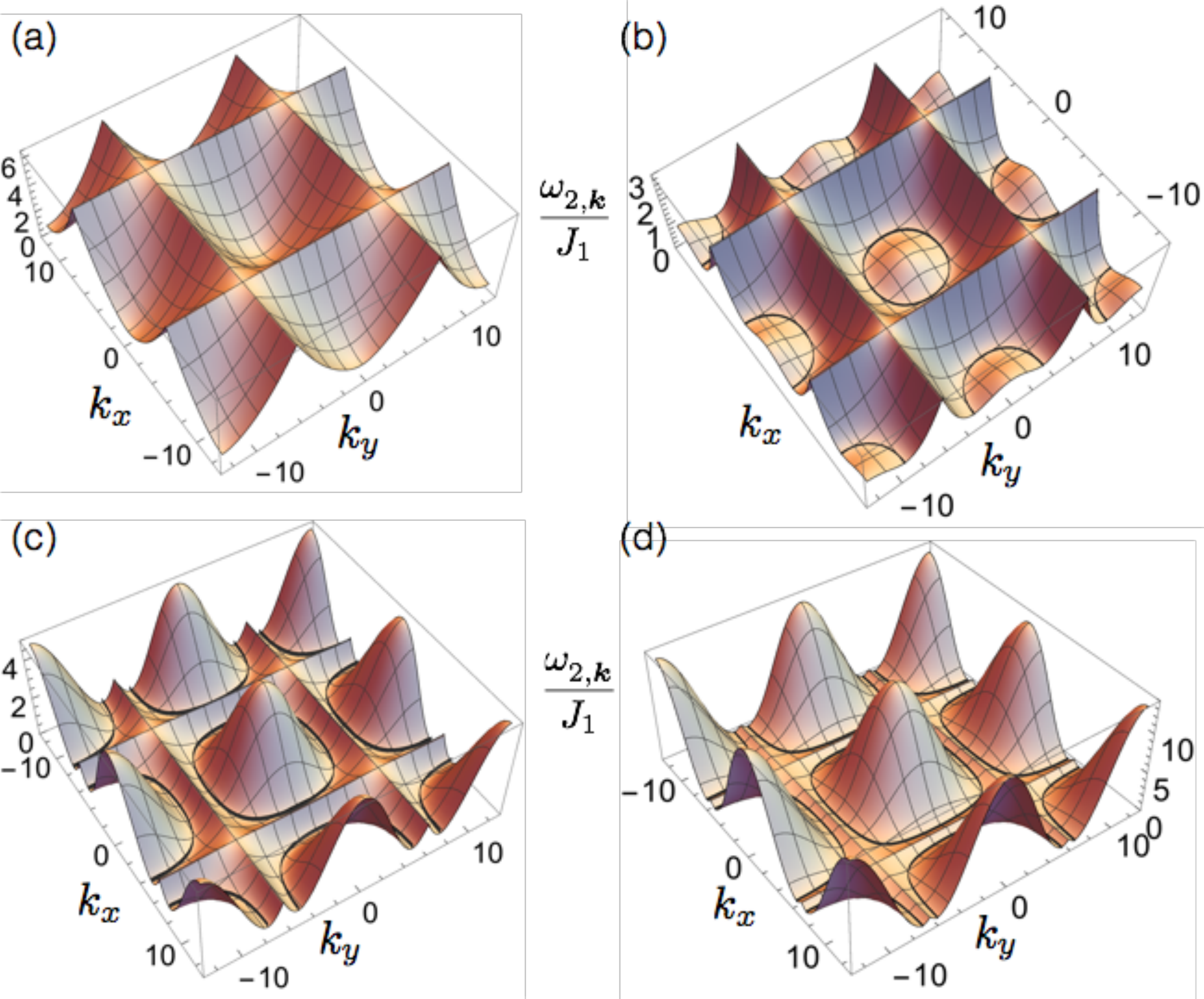}}
\caption{(Color online.) 
The magnetic excitation $\omega_{2,\boldsymbol{k}}$
in the $k_x$-$k_y$ plane of the quantum paramagnet. 
We have chosen the following parameters (a) ${J_2=0.05J_1, D_z = 3J_1}$;
(b) ${J_2 = 0.18J_1, D_z = 1.5 J_1}$;
(c) ${J_2 = 0.4 J_1, D_z = 1.5 J_1}$;
(d) ${J_2 = 0.8 J_1, D_z = 2 J_1}$. 
In the figure, we set ${k_z = 0}$, and an extended zone with ${k_x \in[-4\pi,4\pi], k_y
\in [-4\pi, 4\pi]}$ is used. 
The degenerate minima are marked with contours. 
One can observe the evolution of the band minima.
}
\label{fig3}
\end{figure}

\emph{Frustrated quantum criticality.}---Here we point out the 
nontrivial magnetic excitation in the quantum paramagnetic state
and the resulting frustrated quantum criticality. 
When ${J_2 < J_1/8}$, the band minimum of the lower excitation 
$\omega_{2,{\boldsymbol k}}$ is at the $\Gamma$ point. 
As we increase $J_2$ beyond $J_1/8$, the dispersion minima are 
obtained by minimizing $\xi_{2,{\boldsymbol k}}$. We find that
the minima of $\omega_{2,{\boldsymbol k}}$ are extensively 
degenerate~\cite{PhysRevB.89.201112,PhysRevLett.114.037203} 
and form a two-dimensional surface in the three-dimensional 
reciprocal space that is defined by 
\begin{equation}
\cos \frac{k_x}{2} \cos \frac{k_y}{2} 
+ \cos \frac{k_x}{2} \cos \frac{k_z}{2}
+ \cos \frac{k_y}{2} \cos \frac{k_z}{2}
=  \frac{J_1^2}{16 J_2^2} - 1,  
\label{eq11}
\end{equation}
where we have set the lattice constant to unity. 
This relation coincides with the degenerate spiral surface that
was obtained in the classical treatment of the $J_1$-$J_2$ 
model~\cite{bergman2007order,Simons}. In Fig.~\ref{fig3}, we depict the band 
$\omega_{2,{\boldsymbol k}}$ in the $k_x$-$k_y$ plane with ${k_z = 0}$. 

\begin{figure}[t]
{\includegraphics[width=8cm]{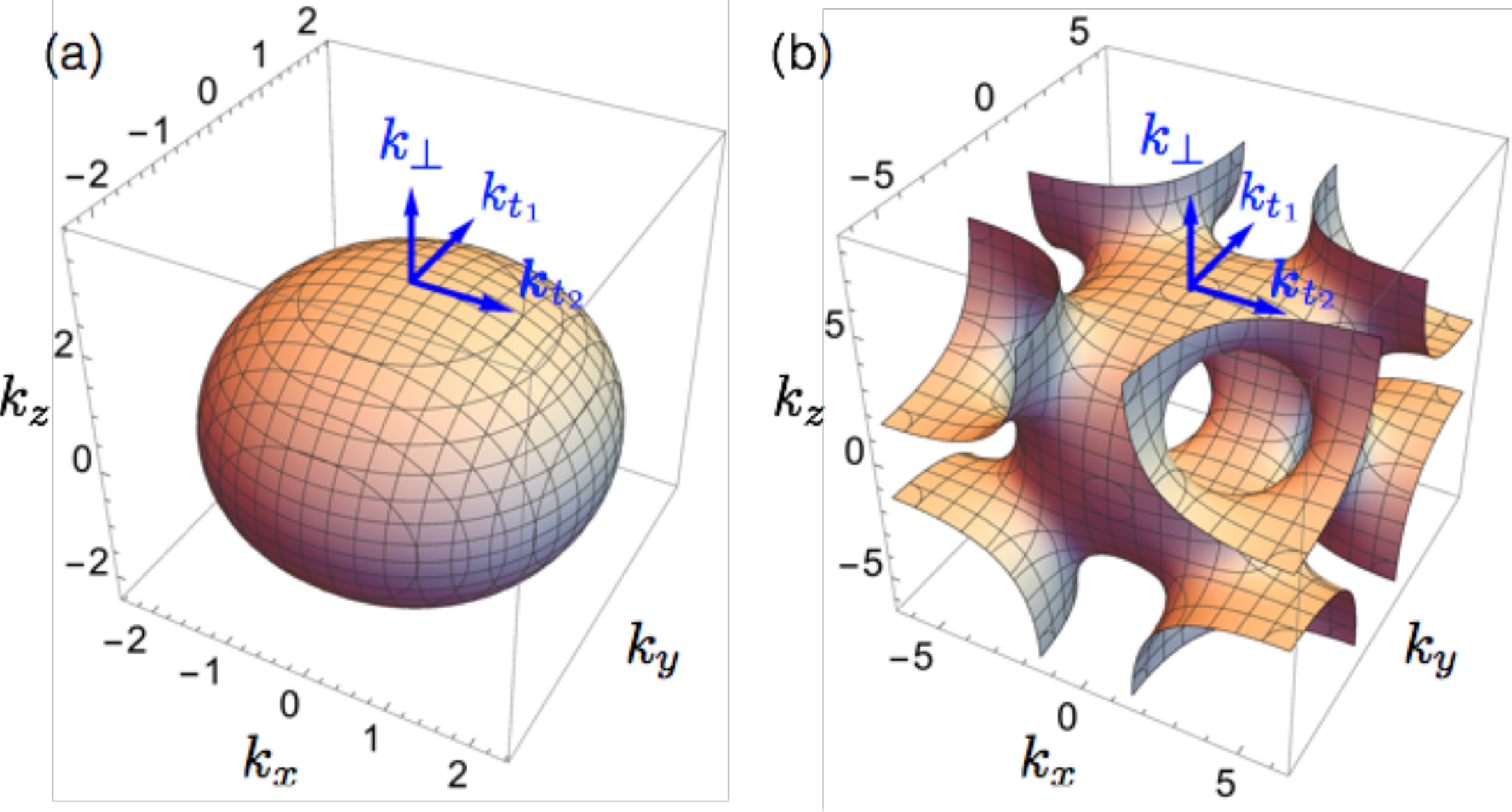}}
\caption{(Color online.) 
The degenerate surface of the band minima at (a) ${J_2 = 0.18 J_1}$
and (b) ${J_{2}= J_1/3}$. The ${(k_{t_1},k_{t_2})}$ are the two tangential
momenta and $k_{\perp}$ is the component normal to the 
degenerate surface. }
\label{fig4}
\end{figure}

Now we explain how the behavior of the heat capacity in the vicinity of
the magnetic critical point are modified by the large density of the 
low-energy excitations near the band minima. For ${J_2 < J_1/8}$, only 
a single bosonic mode becomes critical (see Fig.~\ref{fig3}a) and leads 
to the usual $C_v \propto T^3$ up to a logarithmic correction from the 
quantum fluctuation at the criticality. 
For ${J_2 > J_1/8}$, however, a {\it degenerate surface of 
bosonic modes} become critical at the transition (see Fig.~\ref{fig3}b,c,d). 
To understand the consequence of this unusual phenomena, we 
return to the saddle point equation in Eq.~\eqref{saddle} that 
reduces to 
\begin{eqnarray}
A \int_{0}^{\Lambda} d k_{\perp}^{} \int_{\Sigma} 
d^2{\boldsymbol k}_{t} \frac{ \coth [\frac{\beta}{2} 
(  m^2 + v^2 k_{\perp}^2 )^{\frac{1}{2}} ]}
{ ( m^2 + v^2 k_{\perp}^2 )^{\frac{1}{2}}} 
+ c = 2, \quad\,\,
\label{eqsad}
\end{eqnarray}
where we have singled out the contribution from the critical modes as 
the first term in Eq.~\eqref{eqsad}, $A$ is an unimportant prefactor 
of the integration, and $c$ is approximately $T$-independent
contribution from the remaining part of the excitations. 
In Eq.~\eqref{eqsad}, we have chosen the coordinate 
basis $({\boldsymbol k}_{t},{k_{\perp}})$ such that 
${\boldsymbol k}_{t}\,(k_{\perp})$ refer to the components 
of the momentum tangential to (normal to) the degenerate surface $\Sigma$
(see Fig.~\ref{fig4}),
and $\Lambda$ is the momentum cutoff.
Here the critical mode behaves ${\omega_{2,{\boldsymbol k}} 
\simeq (m^2 + v^2 k_{\perp}^2)^{\frac{1}{2}}}$ 
in which $m$ is the thermally generated mass term and $v$ is the 
velocity normal to the degenerate surface.  
At low temperatures ($T \ll \Lambda$), the temperature dependent 
part of the integral becomes independent of the cutoff $\Lambda$, 
and only depends on $T$ via the dimensionless parameter $m^2/T^2$. 
In order for the equality in Eq.~\eqref{eqsad} to
hold, we expect $m \propto T$.

From the scaling form of $m$, we obtain a remarkable result 
for the low-temperature heat capacity that behaves as $C_v \propto T$
at the criticality. This linear-$T$ heat capacity is like the one in 
a Fermi liquid metal, except that this is a pure bosonic system!  
This unusual behavior simply arises from the frustrated spin 
interaction.

\emph{{Quantum order by disorder.}}---When the extensively degenerate 
modes are condensed at the critical point for ${J_2 > J_1/8}$, 
extensively degenerate candidate ordered states are available, 
and it is the quantum fluctuation of the spins that selects the 
the particular orders in the phase diagram of Fig.~\ref{fig2}.  

To explain this phenomenon, we first realize that the easy-plane spin
anisotropy favors the magnetic order in the $xy$ plane with 
\begin{eqnarray}
{\boldsymbol r} \in \text{A}, \quad {\boldsymbol S}_{\boldsymbol r} & = &  S\, \text{Re} [ (\hat{x} - \mathbb{i} \hat{y}) e^{\mathbb{i}  {\boldsymbol q} \cdot {\boldsymbol r}} ], 
\label{eq14}
\\ 
{\boldsymbol r} \in \text{B}, \quad {\boldsymbol S}_{\boldsymbol r} & = &  S\, \text{Re} [ (\hat{x} - \mathbb{i} \hat{y}) e^{\mathbb{i} {\boldsymbol q} \cdot {\boldsymbol r} + \mathbb{i} 
\theta_{\boldsymbol q} } ],
\label{eq15}
\end{eqnarray}
where ${\boldsymbol q}$ is the propagating wavevector of the 
spin spiral, and $\theta_{\boldsymbol q}$ is the phase shift between A and B sublattices 
of the diamond lattice. Both ${\boldsymbol q}$ and ${\theta}_{\boldsymbol q}$ can be obtained by 
a Weiss mean-field theory that is like
the early classical treatment~\cite{bergman2007order}. 
The quantum fluctuation with
respect to the candidate spin spiral state is analyzed by a linear
spin-wave theory and is discussed in the detail in the Supplementary information. 
As we plot in Fig.~\ref{fig2}, quantum fluctation favors the spiral wavevector 
to be either along [001] or [111] direction. 
For ${J_2 > J_1/4}$, the degenerate surface has expanded to the Brillouin zone  
boundary, and the [111] direction no longer intersects with the degenerate 
surface (see Fig.~\ref{fig4}b as an example),
the six points around the [111] direction are selected, and the resulting ordering
states are labeled by $[111^{\ast}]$ in Fig.~\ref{fig2}.

\emph{Discussion.}---In contrary to the proposal of a topological quantum paramagnet 
in NiRh$_2$O$_4$~\cite{McQueenunPub}, 
our theoretical prediction does not support topological quantum paramagnet in our minimal 
$J_1$-$J_2$-$D_z$ spin model. Instead, due to the strong frustrated spin interaction,
a large region of trivial quantum paramagnet state is stabilized in the phase diagram.
Although the trivial quantum paramagnet does not represent any new state of matter,
the magnetic excitation is rather unusual and supports a degenerate 
surface of band minima in the spectrum. As the system is driven into a magnetic ordered state,
 extensively degenerate critical modes from the degenerate surface 
are condensed, leading to an unconventional critical properties at the transition. 

To differentiate the proposal of topological quantum paramagnet and our 
proposal, we propose the following experiments. 
In a topological quantum paramagnet,
the bulk is fully gapped and the surface may show various anomalous behaviors~\cite{chen2012symmetry,PhysRevX.3.011016,PhysRevB.91.195131,PhysRevX.5.021029}. 
If the system develops gapless surface states, 
it should be detectable by the surface thermal 
transport. If the system realizes intrinsic topological 
order~\cite{PhysRevX.3.011016,PhysRevX.5.021029}, 
one would observe fractionalized excitations on the surface. 
If the system breaks the time reversal symmetry on the surface, then one
would observe a surface magnetic order. In contrast, our prediction 
of the thermodynamic properties and the excitation spectrum
for the trivial quantum paramagnet can be directly measured by 
the bulk measurements such as magnetic susceptibility and inelastic 
neutron scattering. Moreover, since the model is applicable
broadly to spin-one tetragonal diamond materials, it is  
of interest to find similar materials in the spinel families.

Finally, we address the role of other interactions. It has been shown 
{\it classically} for the spin ${S=5/2}$ diamond lattice antiferromagnet
MnSc$_2$S$_4$ that very weak third neighbor interaction could lift
the continuous degeneracy~\cite{bergman2007order}. Here, the 
quantum paramagnetic phase is a robust state, the presence of 
weak further neighbor interaction cannot destabilize it and 
we expect the general structure of the phase diagram in Fig.~\ref{fig2} 
to stay intact. The effect of the other weak interactions
on the excitation in quantum paramagnet is a very low energy scale 
property and may not be visible under the current experimental resolution. 

\emph{Acknowledgements.}---We acknowledge one anonymous referee 
for criticism and comment that helps improve the paper.  
This work is supported by the Ministry of Science and 
Technology of China with the Grant No.2016YFA0301001, 
the Start-Up Funds and the Program of First-Class University 
Construction of Fudan University, and the Thousand-Youth-Talent 
Program of China. We thank Prof Nanlin Wang for the hospitality 
during my visit at ICQM of Peking University, Prof Zhong Wang 
for the hospitality during my visit at IAS of Tsinghua University,
Prof Chen Fang for the hospitality during my visit at IOP,
and Prof Ying Ran for the hospitality during my visit at Boston 
College this January when this work was motivated and finalized. 
We thank Prof Yang Qi and Dr. Peng Ye for a Wechat conversation
that points other surface possibilities that are consistent
with Prof Senthil's later comments. Finally, we thank Prof McQueen 
and Prof Senthil for the recent correspondence and comments. {\sl{Notes added:}}
During the review process of our work, a preprint~\cite{SimonJune} 
appeared and studied a modified exchange model for NiRh$_2$O$_4$.
Their results are complementary to ours.

\bibliography{refs}

\newpage


\begin{widetext}
{\Large Supplementary Information for ``Quantum Paramagnet and Frustrated Quantum Criticality in a Spin-One Diamond Lattice Antiferromagnet''}
\vspace{1cm} 
\\
\noindent{I.} Energy level of orbitals\\
II. The magnetic susceptibility. \\
III. Weiss mean-field theory in the quantum paramagnetic phase. \\
IV. Exchange matrix. \\
V. Quantum order by disorder. \\
\end{widetext}

\section{Energy level of orbitals} 

In the main text, we assume that the degenerate $xz$ and $yz$ orbitals 
are above the $xy$ orbital such that the orbital degree of freedom is 
fully quenched and we obtain the spin-only local moment with $S=1$ for 
the Ni$^{2+}$ ion. Therefore, the atomic spin-orbit coupling is quenched
at the linear order, and we can ignore the effect of the spin-orbit coupling 
if the crystal field splitting within the $t_{2g}$ shell is larger than 
the spin-orbit coupling~\cite{PhysRev.171.466}. Under the above circumstances,
the spin model in the main text is applicable to NiRh$_2$O$_4$. 

\begin{figure}[h]
{\includegraphics[width=5cm]{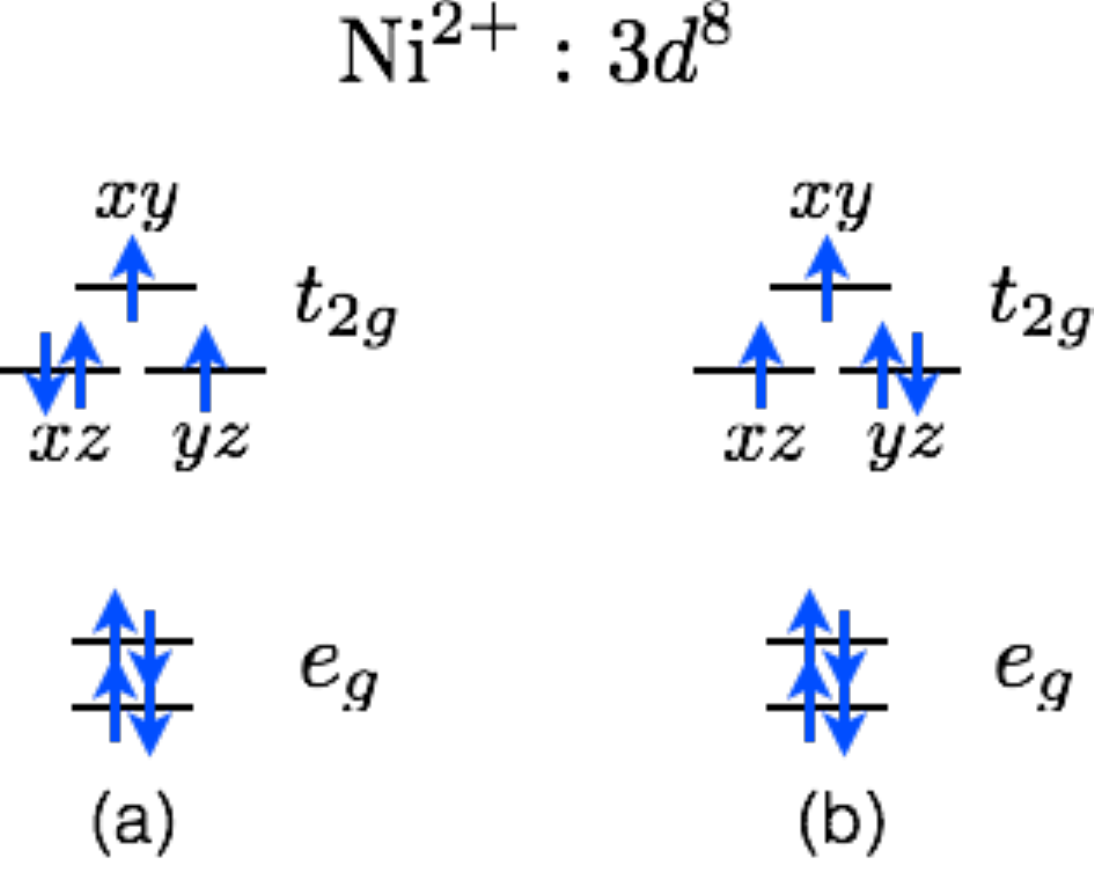}}
\caption{(Color online.) The energy level diagram when the $xz$ and $yz$ orbitals 
are lower than the $xy$ orbital. (a) and (b) are 
two equivalent electron filling schemes and can be represented by a pseudospin-1/2 
orbital degree of freedom. See the text for the detailed discussion. 
}
\label{sfig1}
\end{figure}

If the degenerate $xz$ and $yz$ orbitals are below the $xy$ orbital, 
there are two different electron filling schemes for the eight electrons 
in the Ni$^{2+}$ ion (see Fig.~\ref{sfig1}). This degenerate filling 
arises from the degeneracy of the $xz$ and $yz$ orbitals. 
In this scenario, a pseudospin-1/2 degree of freedom would be introduced to 
describe the two-fold orbital degeneracy. Since the $xz$ and $yz$ orbitals 
belong to the $t_{2g}$ orbital, the atomic spin-orbit coupling is active at the 
linear order~\cite{WCKB}. Due to the orbital degree of freedom, the exchange 
interaction would be a Kugel-Khomskii superexchange interaction~\cite{Kugel82}. 
The potential interplay between the {\it on-site atomic spin-orbit coupling} and 
the {\it inter-site Kugel-Khomskii superexchange interaction} will be discussed in 
the future work. 

It is not quite obvious right now whether the energy 
level scheme in the main text or in Fig.~\ref{sfig1} applies to 
NiRh$_2$O$_4$. To resolve them, it would be nice to carry out an optical 
measurement to detect the energy level scheme of the Ni$^{2+}$ ion. 
The existing experiment in Ref.~\onlinecite{McQueen} used ZnRh$_2$O$_4$ 
to subtract the phonon contribution to the specific heat for NiRh$_2$O$_4$
and found the magnetic entropy exceeds the spin-only contribution. NiRh$_2$O$_4$
experiences a structural transition at 380K, and it is not obvious whether
ZnRh$_2$O$_4$ has the same structure as NiRh$_2$O$_4$. So using ZnRh$_2$O$_4$
as the phonon background may be debatable. 

\section{The magnetic susceptibility}

The presence of the single-ion anisotropy modifies the Curie-Weiss temperature. 
Since $D_z > 0$ corresponds to the easy-plane spin anisotropy and tends to orients 
the spin in the $xy$ plane, so we expect $D_z$ to contribute an antiferromagnetic (ferromagnetic)
Curie-Weiss temperature when the external field is applied along the $z$ direction 
(in the $xy$ plane). To examine this, here we carry out the high temperature expansion and extract 
the Curie-Weiss temperature. In a magnetic field that is applied along the $z$ direction, 
the Hamiltonian is 
\begin{eqnarray}
H_{h} = \sum_{{\boldsymbol r}, {\boldsymbol r}'} J_{{\boldsymbol r} {\boldsymbol r}'}
{{\boldsymbol S}_{\boldsymbol r} \cdot {\boldsymbol S}_{{\boldsymbol r}'} }
+ \sum_{\boldsymbol r} 
\big[ D_z (S^z_{\boldsymbol r})^2 - h \, S^z_{\boldsymbol r} \big].
\end{eqnarray}
The magnetization $M^z$ is then given as 
\begin{eqnarray}
M^z = \sum_{\boldsymbol r} 
\frac{\text{Tr}[ S_{\boldsymbol r}^z e^{-\beta H_{h}} ]}
{\text{Tr} [ e^{-\beta H_{h}}]},
\end{eqnarray}
from which one can carry out the expansion in $\beta = 1/(k_{\text B} T)$. 
The linear order in $\beta$ is the free ion contribution without the single-ion
anisotropy $D_z$. For the second order terms in $\beta$,  besides the usual contribution from the 
crossing term betweeen the superexchange and the Zeeman coupling, we now
have a new contribution from the crossing term between the single-ion anisotropy
and the Zeeman coupling. These crossing terms make non-vanishing contributions to the
magnetization. From the magnetization, it is straightforward to read the susceptibility
and the Curie-Weiss temperature. Likewise, the magnetization and the Curie-Weiss 
temperature for the field in the $xy$ plane can also be obtained.

\section{Weiss mean-field theory in the quantum paramagnet}

Here we explain the zero-temperature spin susceptibility in the quantum paramagnet.
From a general symmetry point of view, the global U(1) spin rotational symmetry around
the $z$ direction is preserved in the quantum paramagnet, so the total $S^z$ is a good
quantum number and we can use the total $S^z$ to label all the states. 
The quantum paramagnet is a {\it gapped} spin singlet state with 
${\sum_{\boldsymbol r} S^z_{\boldsymbol r} = 0}$. Thus it is obvious that the spin 
susceptibility for field applied along $z$ direction is zero. 

For the magnetic field in the $xy$ plane, the system no longer has a global U(1) 
symmetry, and the above argument fails. To calculate the zero-temperature spin 
susceptibility, we take a Weiss mean-field approach and replace the 
spin model with a mean-field model, {\it i.e.},
\begin{eqnarray}
&& H_x^{} = \sum_{{\boldsymbol r}{\boldsymbol r}'} 
J_{{\boldsymbol r}{\boldsymbol r}'}^{} 
{\boldsymbol S}_{\boldsymbol r} \cdot 
{\boldsymbol S}_{{\boldsymbol r}'} 
+  \sum_{\boldsymbol r} \big[ D_z^{} (S^z_{\boldsymbol r})^2 
-  h_x^{} S^x_{\boldsymbol r} \big] \quad
\\
&& \quad\quad\quad\quad\quad\quad\quad\quad\quad\quad \Downarrow 
\nonumber 
\\
&& H_{\text{MFx}}^{} =
\sum_{{\boldsymbol r}{\boldsymbol r}'} 
J_{{\boldsymbol r}{\boldsymbol r}'}^{} 
{S}^x_{\boldsymbol r}  \langle {S}^x_{{\boldsymbol r}'} \rangle 
+  \sum_{\boldsymbol r} \big[ D_z^{} (S^z_{\boldsymbol r})^2 
-  h_x^{} S^x_{\boldsymbol r} \big] \quad,
\end{eqnarray}
where $\langle S^x_{\boldsymbol r} \rangle \equiv m^x $ and we assume
a uniform mean-field ansatz. We solve for $m^x$ self-consistently 
and obtain the magnetization,
\begin{eqnarray}
m^x  = \frac{2 h_x}{D_z + 2(z_1 J_1 + z_2 J_2)}
\end{eqnarray}
and the spin susceptibility
\begin{equation}
\chi_{\perp}^{} = \frac{2\mu_0 (g \mu_{\text B})^2}{D_z + 2 (z_1 J_1 + z_2 J_2)},
\end{equation}
where we have put back in the physical units.

\section{Exchange matrix}

The exchange matrix, that was introduced in the main text, is simply 
obtained by Fourier transform of the exchange part of the Hamiltonian.
We have
\begin{eqnarray}
{\mathcal J}_{\boldsymbol k} = \left[ 
\begin{array}{ll}
J_2 \sum_{\mu=1}^{12} e^{\mathbb{i} {\boldsymbol k}\cdot {\boldsymbol b}_{\mu} } & 
J_1 \sum_{\mu=1}^4 e^{\mathbb{i} {\boldsymbol k}\cdot {\boldsymbol a}_{\mu} } 
\vspace{2mm}
\\
J_1 \sum_{\mu=1}^4 e^{-\mathbb{i} {\boldsymbol k}\cdot {\boldsymbol a}_{\mu} } & 
J_2 \sum_{\mu=1}^{12} e^{\mathbb{i} {\boldsymbol k}\cdot {\boldsymbol b}_{\mu} }
\end{array}
\right],
\end{eqnarray}
where ${\boldsymbol a}_{\mu}$ are the four first neighbor vectors and ${\boldsymbol b}_{\mu}$
are the twelve second neighbor vectors. 

The eigenvalues of ${\mathcal J}_{\boldsymbol k}$ are easily obtained
\begin{eqnarray}
\xi_{1,{\boldsymbol k}} &=& 4J_2 \alpha_{\boldsymbol k} + 2 J_1 (1 + \alpha_{\boldsymbol k})^{1/2} ,
\\ 
\xi_{2,{\boldsymbol k}} &=& 4J_2 \alpha_{\boldsymbol k} - 2 J_1 (1 + \alpha_{\boldsymbol k})^{1/2} ,
\end{eqnarray}
where
\begin{eqnarray}
\alpha_{\boldsymbol k} =\cos \frac{k_x}{2} \cos \frac{k_y}{2} + 
\cos \frac{k_x}{2} \cos \frac{k_y}{2} + \cos \frac{k_x}{2} \cos \frac{k_y}{2}. \quad 
\end{eqnarray}

\section{Quantum order by disorder}

In the ordered regime, the system develops a spin spiral order in the $xy$ plane,
and the mean-field theory for the ordered state yields the mean-field Hamiltonian 
for the $xy$ spin components, 
\begin{eqnarray}
H_{xy} =\frac{1}{2} \sum_{\boldsymbol q} \sum_{i,j} 
{\mathcal J}_{{\boldsymbol q},ij}
({S}^x_{i,{\boldsymbol q}}   
{S}^x_{j,-{\boldsymbol q}} +
{S}^y_{i,{\boldsymbol q}}   
{S}^y_{j,-{\boldsymbol q}} ).
\end{eqnarray}
The ordering wavevector ${\boldsymbol q}$ and the phase shift 
$\theta_{\boldsymbol q}$ are determined by optimizing the eigenvalue 
and corresponding eigenvector of the exchange matrix. 
The optimal ${\boldsymbol q}$ satisfies Eq.11 in the main text
and forms a degnerate surface when ${J_2 > J_1/8}$, 
and this result is identical to the early classical 
treatment in Ref.~\onlinecite{bergman2007order}. For the spin spiral 
state that is defined in Eq.13 and Eq.14 of the main text, 
the combined operation of the lattice translation and spin rotation around 
the $z$ axis by the spiral angle remains to be a symmetry. We thus 
introduce the following Holstein-Primakoff boson for the spin spiral 
state,
\begin{eqnarray}
&& {\boldsymbol S}_{\boldsymbol r} \cdot \hat{n}_{\boldsymbol r}= S - a^{\dagger}_{\boldsymbol r}a^{}_{\boldsymbol r},
\\
&&{\boldsymbol S}_{\boldsymbol r} \cdot \hat{z} = \frac{\sqrt{2S}}{2} 
( a^{}_{\boldsymbol r} + a^{\dagger}_{\boldsymbol r}  ) ,
\\
&& {\boldsymbol S}_{\boldsymbol r} \cdot ( \hat{n}_{\boldsymbol r}\times \hat{z} ) = \frac{\sqrt{2S}}{2\mathbb{i}}
(a^{}_{\boldsymbol r} - a^{\dagger}_{\boldsymbol r} ),
\end{eqnarray} 
where $\hat{n}_{\boldsymbol r}$ is the orientation of the spin spiral order at the lattice
site ${\boldsymbol r}$. With this substitution of the spin operators, we obtain 
the linear spin-wave Hamiltonian that is given as
\begin{eqnarray}
H_{\text{sw}} &=& \sum_{{\boldsymbol k}\in {\text{BZ}}} 
(a^{\dagger}_{1{\boldsymbol k}},a^{\dagger}_{2{\boldsymbol k}},
a^{}_{1,-{\boldsymbol k}}, a^{}_{2,-{\boldsymbol k}}) \nonumber \\
&\times &  \left[
\begin{array}{llll}
A_{{\boldsymbol k},11} & A_{{\boldsymbol k},12}  & B_{{\boldsymbol k},11} & B_{{\boldsymbol k},12} \\
A_{{\boldsymbol k},12}^{\ast} & A_{{\boldsymbol k},22} & B_{-{\boldsymbol k},12}& B_{{\boldsymbol k},22} \\ 
B_{{\boldsymbol k},11}^{\ast} & B_{-{\boldsymbol k},12}^{\ast} & A_{-{\boldsymbol k},11} 
& A^{\ast}_{-{\boldsymbol k},12} \\
B_{{\boldsymbol k},12}^{\ast} & B_{{\boldsymbol k},22}^{\ast} & A_{-{\boldsymbol k},12} & A_{-{\boldsymbol k},22} 
\end{array}
\right]
\left(
\begin{array}{l}
a^{}_{1,{\boldsymbol k}} \\
a^{}_{2,{\boldsymbol k}} \\
a^{\dagger}_{1,-{\boldsymbol k}} \\
a^{\dagger}_{2,-{\boldsymbol k}} 
\end{array}
\right)
\nonumber \\
&-& \sum_{{\boldsymbol k}\in {\text{BZ}}} (A_{{\boldsymbol k},11 } + A_{{\boldsymbol k},22 }),
\end{eqnarray}
where 
\begin{eqnarray}
&&A_{{\boldsymbol k},11} = A_{{\boldsymbol k},22} 
= \frac{D_z}{2} - \frac{J_1}{2} \sum_{\mu=1}^4 \cos({\boldsymbol q} \cdot {\boldsymbol a}_{\mu} 
+ \theta_{\boldsymbol q})  
\nonumber \\
&& \, + \frac{J_2}{4} \sum_{\mu=1}^{12} \big[ { \cos ({\boldsymbol k}\cdot {\boldsymbol b}_{\mu})
+ \big( {\cos ({\boldsymbol k}\cdot {\boldsymbol b}_{\mu}) - 2 } \big) 
\cos ({\boldsymbol q} \cdot {\boldsymbol b}_{\mu} ) } \big] , \nonumber \\
\\
&& A_{{\boldsymbol k},12} = \frac{J_1}{4} \sum_{\mu=1}^4 e^{\mathbb{i} {\boldsymbol k} \cdot {\boldsymbol a}_{\mu}} [1 + \cos ({\boldsymbol q}\cdot {\boldsymbol a}_{\mu} + \theta_{\boldsymbol q}) ] , \\
&& B_{{\boldsymbol k},11} =\frac{D_z}{2} +  \frac{J_2}{4} \sum_{\mu=1}^{12} 
   \cos ( {\boldsymbol k} \cdot {\boldsymbol b}_{\mu} ) [1- \cos ({\boldsymbol q}\cdot {\boldsymbol b}_{\mu}) ], \\
&& B_{{\boldsymbol k},12} = \frac{J_1}{8} \sum_{\mu=1}^4 
\cos ({\boldsymbol k} \cdot {\boldsymbol a}_{\mu})   
 [ 1 - \cos ({\boldsymbol q}\cdot {\boldsymbol a}_{\mu} + \theta_{\boldsymbol q} )],
\end{eqnarray}
and $a_{1,{\boldsymbol k}}$ and $a_{2,{\boldsymbol k}}$ represent the 
Holstein-Primakoff boson on the A and the B sublattices, respectively. 

The spin-wave Hamiltonian is diagonalized by a Bogoliubov transformation. 
The quantum zero point energy is given as 
\begin{eqnarray}
\Delta E = \sum_{{\boldsymbol k}}\sum_{i=1}^2 \frac{1}{2} \Omega_{{\boldsymbol k},i} -
\sum_{\boldsymbol k} 2 A_{{\boldsymbol k},11} ,
\end{eqnarray}
where $\Omega_{{\boldsymbol k},i}$ is the $i$-th spin-wave mode 
at the momentum ${\boldsymbol k}$. 

In the phase diagram in Fig.2 of the main text, the propagating wavevectors of 
the [111] spin spiral and the [001] spin spiral are uniquely specified 
by the intersection between the orientation and the degenerate
surface. Here we describe the [111$^{\ast}$] spin spirals.  
As we have already pointed out in the main text, for ${J_2 > J_1/4}$, there is
no intersection beween the 111 axis and the degenerate surface. Instead, the 
Brillouin zone boundary/surface, that is normal to the 111 axis, intersects with 
the degenerate surface, and the interaction is a deformed circle (see Fig.4b). 
The quantum fluctuation selects the propagating wavevector on this deformed circle.
Due to the cubic symmetry, six equivalent wavevectors on the deformed circle are selected.

\end{document}